\documentclass[fleqn,usenatbib]{mnras}

\usepackage{newtxtext,newtxmath}
\usepackage[T1]{fontenc}
\usepackage{ulem}

\DeclareRobustCommand{\VAN}[3]{#2}
\let\VANthebibliography\thebibliography
\def\thebibliography{\DeclareRobustCommand{\VAN}[3]{##3}\VANthebibliography}

\usepackage{graphicx}	% Including figure files
\usepackage{amsmath}	% Advanced maths commands
\usepackage[dvipsnames]{xcolor}

\title[Cosmology with clusters and voids]{Exploring the cosmological synergy between galaxy cluster and cosmic void number counts}

\author[D. Pelliciari et al.]{
D. Pelliciari,$^{1,2}$\thanks{E-mail: davide.pelliciari@inaf.it}
S. Contarini,$^{1,3,4}$
F. Marulli,$^{1,3,4}$
L. Moscardini,$^{1,3,4}$
C. Giocoli,$^{3,4}$
G. F. Lesci,$^{1,3}$
\newauthor
and K. Dolag $^{6,7}$
\\
\\
$^{1}$ Dipartimento di Fisica e Astronomia "Augusto Righi" - Alma Mater
Studiorum Università di Bologna, via Piero Gobetti 93/2, I-40129
Bologna, Italy\\
$^{2}$ Istituto Nazionale di Astrofisica (INAF) – Istituto di Radioastronomia (IRA), via Gobetti 101, 40129 Bologna, Italy\\
$^{3}$ Istituto Nazionale di Astrofisica (INAF) - Osservatorio di Astrofisica e Scienza dello Spazio (OAS), via Piero Gobetti 93/3, I-40129 Bologna, Italy\\
$^{4}$ Istituto Nazionale di Fisica Nucleare (INFN) - Sezione di Bologna, viale Berti Pichat 6/2, I-40127 Bologna, Italy\\
$^{6}$ Universitats-Sternwarte, Fakultat fur Physik, Ludwig-Maximilians-Universitat Munchen, Scheinerstr.1, 81679 Munchen, Germany\\
$^{7}$ Max-Planck-Institut fur Astrophysik, Karl-Scharzschildstr. 1, D-85748, Garching, Germany
}

\date{Accepted XXX. Received YYY; in original form ZZZ}

\pubyear{2022}

\begin{document}
\label{firstpage}
\pagerange{\pageref{firstpage}--\pageref{lastpage}}
\maketitle

\begin{abstract}
Galaxy clusters and cosmic voids, the most extreme objects of our Universe in terms of mass and size, trace two opposite sides of the large-scale matter density field. By studying their abundance as a function of their mass and radius, respectively, i.e. the halo mass function (HMF) and void size function (VSF), it is possible to achieve fundamental constraints on the cosmological model. While the HMF has already been extensively exploited providing robust constraints on the main cosmological model parameters (e.g. $\Omega_{\rm m}$, $\sigma_8$ and $S_8$), the VSF is still emerging as a viable and effective cosmological probe. Given the expected complementarity of these statistics, in this work we aim at estimating the costraining power deriving from their combination. To this end, we exploit realistic mock samples of galaxy clusters and voids extracted from state-of-the-art large hydrodynamical simulations, in the redshift range $0.2 \leq z \leq 1$.  We perform an accurate calibration of the free parameters of the HMF and VSF models, needed to take into account the differences between the types of mass tracers used in this work and those considered in previous literature analyses. Then, we obtain constraints on $\Omega_{\rm m}$ and $\sigma_8$ by performing a Bayesian analysis. We find that cluster and void counts represent powerful independent and complementary probes to test the cosmological framework. In particular, the constraining power of the HMF on $\Omega_{\rm m}$ and $\sigma_8$ improves drastically with the VSF contribution, increasing the $S_8$ constraint precision by a factor of about $60\%$.
\end{abstract}


%%%%%%%%%%%%%%%%%%%%%%%%%%%%%%%%%%%%%%%%%%%%%%%%%%

%%%%%%%%%%%%%%%%% BODY OF PAPER %%%%%%%%%%%%%%%%%%

\section{Introduction}
Thanks to the remarkable achievements obtained by experiments on the cosmological microwave background \citep[CMB, see e.g.][]{WMAP7, Planck2018} as well as by spectroscopic and photometric galaxy surveys \citep[e.g.][]{BOSS_alam2017}, cosmology has recently entered a new \textit{precision} era, in which cosmological parameters are constrained with sub-percent accuracy. The $\Lambda$-cold dark matter ($\Lambda$CDM) model, being described by only six parameters, got strong empirical support by the current cosmological observations. In this framework, the matter-energy budget of our Universe is dominated by two dark components, i.e. dark energy (DE), responsible for its present accelerated expansion \citep{Riess1998, Perlmutter1999}, and cold dark matter (CDM), which is the dominant matter component in galaxies and galaxy clusters.

However, statistically significant tensions between the cosmological constraints derived by early-and late-time measurements have also arisen in the last years \citep[see e.g.][for recent reviews]{CI12020, CI32020, CI42020, Perivolaropoulous21}. Among those, the most relevant ones are the discrepancies between the derived values of the Hubble constant, $H_0$, and the tension on the growth of cosmic structures, often quantified in terms of the parameter $S_8 \equiv \sigma_8 \sqrt{\Omega_{\rm m}/0.3}$, where $\Omega_{\rm m}$ is the matter density parameter and $\sigma_8$ is related to the amplitude of the matter power spectrum. Many works have been focused on these tensions \citep[see e.g.][for a recent review of astrophysical and cosmological tensions]{Abdalla22}. In recent years numerous theoretical solutions have been proposed, going from modified gravity theories to alternative DE theories \citep[see e.g.][for an extensive review]{YooWatanabe12, Amendola13, Joyce16}.

To shed light on these tensions it is fundamental to consider different and complementary cosmological probes, whose synergy can improve the precision on the parameter measurements. In fact, the highest precision is achieved by considering probe combination, i.e. computing the joint constraints from different analyses, such as e.g. combining CMB anisotropies and large-scale strucures observations \citep{Webster98, Gawiser_Silk, Bridle99}, as well as supernovae Type-Ia (SNIa), baryonic acoustic oscillations, weak gravitational lensing and galaxy clustering \citep[see e.g.][]{DES2019}. Moreover, \citet{Wu22} recently discussed the constraining power of the combination between 21 cm intensity mapping, fast radio bursts, gravitational wave standard sirens and strong gravitational lensing, which are novel probes that will be greatly developed in the near future \citep[see][for a recent review about alternative cosmological probes]{Moresco22}. 

In terms of mass and size, galaxy clusters and cosmic voids are the most rare and extreme objects in the Universe. Very massive clusters and large voids follow the exponential tail of their mass/size distribution functions, which makes their abundance a probe very sensitive to the change of cosmological parameters. It is well known that galaxy clusters, associated with massive DM haloes, provide a powerful cosmological probe when exploiting their abundances \citep{Vikhlinin09, Cunha09, Balaguera13, Hasselfield13, PlanckCC, PlanckClusterCounts, Bouquet2019, Costanzi2019, Lesci2022a}, as well as their clustering properties \citep{Hong12, Veropalumbo16, Marulli2018, Marulli2021, Lesci2022b}. Also, cosmological constraints become even more stringent when cluster counts are combined with clustering measurements \citep[see e.g.][]{Mana13, salvati18}, 

Concerning cosmic voids, their number counts \citep{Platen08, Pisani15, Contarini2019, Contarini22Euc}, density profiles \citep{Hamaus14, Ricciardelli14, Nadathur15}, and redshift-space distortions of void-galaxy cross-correlation \citep{Nadathur20, Hamaus2020, Hamaus2022}, are rapidly becoming competitive and alternative cosmological probes. Indeed, their unique low density interiors and large sizes make these objects powerful tools to study the properties of the DE \citep{Bos, Pisani15, Verza19} and the effects of modified gravity theories \citep{Spolyar, Barreira, Contarini2020}, to which they are highly sensitive. Cosmic voids are also well suited to test primordial non-Gaussianity \citep{chan2019}, and other physical phenomena beyond the Standard Model of particle physics \citep{Reed2015, yang2015, baldi2016}, and to investigate elusive components such as massive neutrinos \citep{Villaescusa-Navarro, massara2015, Shuster, kreisch2019, Contarini2020}, which are predominant in these regions.

The number counts of galaxy clusters and cosmic voids can be considered as independent and complementary cosmological probes given the opposite regimes in sides of the matter density field they map. We then expect some degrees of orthogonality between the corresponding cosmological constraints \citep{Sahlen16, Bayer21, Kreisch21}. This important feature represents a desired property to extract as much information as possible from the joint combination of these two probes, and it has not been exploited in great details so far.

In this work we investigate the constraining power of a joint analysis of the number counts of galaxy clusters and cosmic voids. We consider redshift-space cluster and void catalogues extracted from a large hydrodynamic cosmological simulation in four redshift bins, spanning in the range $0.2 \leq z \leq 1$. Recent works have already studied the constraining power derived from the combination of halo and void statistics obtained from cosmological simulations \citep{Bayer21, Kreisch21}. Nevertheless, the analysis we present in this work is directly applicable to real data sets, thanks to the bias-dependent void size function parametrisation (Section \ref{VSFmodel}), the inclusion of the most relevant observational effects (Section \ref{sec:obs_eff}), the model calibrations on hydrodynamical simulations (Section \ref{sec:calibrationHMF}), and the both realistic and conserving mass and radius selections considered for clusters and voids. 

The paper is structured as follows. In Section \ref{models} we introduce the theoretical context of the analysis we carry out in this work, focusing on the considered halo mass function and void size function models. In Section \ref{MAGNETICUM} we describe the simulations used to construct cluster and void mock catalogues, along with the prescriptions for a reliable data set preparation. In Section \ref{methods} we introduce the observational effects considered in the simulation analysis, and we present the Bayesian framework used to assess the posterior distributions of the model parameters. We also estimate the correlation between the two cosmological probes considered, verifying their statistical independence. Subsequently, we outline in Section \ref{results} the adopted calibration method considered for the analysis, required for an accurate comparison between the data and the theoretical models, and we show the results of the combined analysis of cluster and void number counts. Finally, we draw our conclusions in Section \ref{conclusions}.

\section{Cosmological probes}\label{models}
The abundance of both haloes and cosmic voids can be modelled with the excursion-set formalism \citep{Bond1991, Zentner2007}, also known as the extended Press-Schechter formalism \citep{PS74}, which can be used to describe the evolution of spherical overdensities and underdensities. In particular, haloes form from the growth of positive fluctuations, i.e. overdensities, while cosmic voids form from the evolution of negative ones, i.e. underdensities. In what follows, we describe the two theoretical models used in this work, i.e. the halo mass function (HMF) and the void size function (VSF), which describe the comoving number density of haloes and voids, as a function of their mass and radius, respectively. We then compute number counts from number density by taking into account the most relevant observational effects (see Section \ref{sec:obs_eff}).

\subsection{Cluster counts}\label{sec:clustercounts}
Following the Press-Schechter formalism, the comoving number density, $n(M,z)$, of DM haloes having masses between $M$ and $M + \mathrm{d}M$ at redshift $z$ can be written as \citep{PS74, Cole1991, Bond1991, Mo&White1996, ST99, ST01}:
\begin{equation}
   \frac{\mathrm{d}n(M, z)}{\mathrm{d}\ln M} = \frac{\bar{\rho}}{M} f_{\ln\sigma}(M, z) \frac{\mathrm{d} \ln \sigma^{-1}(M,z)}{\mathrm{d} \ln M} \textrm{ ,}
\end{equation}
when expressed through its logarithmic differential. Here $\bar{\rho}$ is the mean cosmic background density
and $f_{\ln\sigma}(M, z)$ is the so-called \textit{multiplicity function}, which depends on the mass definition adopted to identify simulated haloes \citep[see e.g.][]{Tinker08}. For the Press-Schechter mass function, the latter is defined as:
\begin{equation}
    f_{\ln \sigma}(M, z) = \frac{\mathrm{d}\sigma(M,z)}{\mathrm{d}\ln M} =  \sqrt{\frac{2}{\pi}}\frac{\delta_{\rm c}}{\sigma_M}\exp\Bigl(-\frac{\delta_{\rm c}^2}{2\sigma^2(M,z)}\Bigr) \textrm{ ,}
\end{equation}
where $\delta_{\rm c}$ is the critical overdensity threshold computed in the linear theory. Here $\sigma_M$ is the square root of the variance of the linear density field extrapolated to the redshift at which haloes are identified, after smoothing with a spherical top-hat filter with a radius $R$ enclosing a mass $M$. This can be expressed in terms of the present-day matter power spectrum $P(k)$:
\begin{equation}\label{variance}
\sigma^2(M,z) = \frac{D^2(z)}{2\pi^3} \int \mathrm{d}^3 \vec{k} P(\vec{k}, z=0) \hat{W}^2[\vec{k}; R(M)] \textrm{ ,}
\end{equation}
where $D(z)$ is the growth factor of linear perturbations, normalised
such that $D(z=0) = 1$ \citep{CPT92}, $\hat{W}[\vec{k}; R(M)]$ is the
Fourier-transform of the window function and $R(M)$ is obtained by requiring that $M/(4\pi R^3 /3)=\bar{\rho}$.
Here $\bar{\rho}$ is expressed in comoving coordinates, so its value is independent of the redshift. The dependence of the HMF on the cosmological parameters $\Omega_{\rm m}$ and $\sigma_8$ lies both in the mass variance, $\sigma^2(M,z)$, and in the mean density of the Universe, $\bar{\rho}$.

It can be shown that for a $\Lambda$CDM cosmology the HMF assumes a universal functional form when the mass variance is parametrised by means of the following scaled variable \citep{ST99, ST01, Courtin11, Despali}:
\begin{equation}
    \nu \equiv \frac{\delta^2_c}{\sigma_M^2} \textrm{ .}
\end{equation}
With this parametrisation the HMF can be written as:
\begin{equation}
    \nu f(\nu) = \frac{M^2}{\bar{\rho}} \frac{\mathrm{d}n}{\mathrm{d}M}\frac{\mathrm{d}\ln M}{\mathrm{d}\ln \nu} \textrm{ ,}
\end{equation}
where:
\begin{equation}\label{HMFmodel}
    \nu f(\nu) = A \biggl(1 + \frac{1}{\nu'^{p}}\biggr)
                   \biggl( \frac{\nu'}{2\pi} \biggr)^{1/2} 
                   e^{-\nu'/2} \textrm{ ,}
\end{equation}
with $\nu' = a \nu$. The parameters of this model are $(a, p, A)$ and can be calibrated using cosmological simulations \citep{Manera10, Courtin11}. In particular, $a$, $p$, and $A$ define the mass function cut-off at high masses, its shape in the low mass range and its normalisation, respectively. 
\citet{Despali} found that the HMF parametrisation of Eq. \eqref{HMFmodel} is universal if clusters are defined by means of the virial overdensity $\Delta_{\rm vir}$. However, since both observed and simulated clusters are usually defined via other overdensity criteria, an extended universal fitting formula was provided by \citet{Despali} as follows:
\begin{equation}\label{fitting}
\begin{aligned}
a &= 0.4332x^2 + 0.2263x + 0.7665\\
p &= -0.1151x^2 + 0.2554x + 0.2488\\
A &= -0.1362x + 0.3292 \textrm{ ,}
\end{aligned}
\end{equation}
where the quantity $x \equiv \log[\Delta(z)/\Delta_\text{vir}(z)]$ properly rescales the spherical overdensity used to define the clusters, $\Delta(z)$, with the virial overdensity, $\Delta_\text{vir}(z)$.

The HMF universality highlighted by \citet{Despali} is a fundamental feature which can be exploited in cosmological analyses. Indeed, its universal shape can be used to derive unbiased cosmological constraints by modelling data at any redshift and for \textit{standard} cosmological scenarios. In particular, the coefficients in Eq. \eqref{fitting} have been calibrated considering DM-only cosmological simulations, with haloes identified with a spherical overdensity halo finder \citep[e.g.][]{Tormen98, Tormen04, Giocoli08}. A re-calibration of the fitting relation parameters is therefore necessary when using different type of simulations (e.g. hydrodynamic) and/or other halo finder methods (see Section \ref{calibrationHMF}).

\subsection{Void counts}\label{VSFmodel}
As already mentioned, the abundance of cosmic voids as a function of their radius can be modelled as well with the excursion-set formalism \citep{SvdW04, Jennings}. Given the analogies among overdensity and underdensity, the linear VSF can be written as:
\begin{equation}\label{VSF_L}
  \frac{\mathrm{d}n}{\mathrm{d}\ln R}\biggr|_{\rm L} =\frac{f_{\ln\sigma}}{V(R_{\rm L})} \frac{\mathrm{d} \ln \sigma^{-1}(R,z)}{\mathrm{d} \ln R_{\rm L}} \mathrm{ ,}
\end{equation}
where $V(R_{\rm L}) = \frac{4}{3}\pi R_{\rm L}^3$ is the volume of a spherical region of radius $R_{\rm L}$ and the subscript "L" indicates that the quantities are obtained in linear theory. $\sigma^2(R,z)$ is the variance of the matter power spectrum on a scale $R$ and, analogously to $\sigma^2(M,z)$ entering in the HMF, embeds the cosmological dependence of the model. Here the multiplicity function can be expressed as an infinite series \citep{SvdW04}:

\begin{equation}\label{flnsigma}
  f_{\ln\sigma} = 2 \sum_{j=1}^{\infty}j \pi x^2 \sin(j \pi
  \mathcal{D})\exp\biggl[-\frac{(j \pi x)^2}{2}\biggr] \textrm{ ,}
\end{equation}
where:
\begin{equation}
    \mathcal{D} \equiv \frac{|\delta_{\rm v, L}|}{\delta_{\rm c} + |\delta_{\rm v, L}|} \quad \textrm{,} \quad x \equiv \frac{\mathcal{D}}{|\delta_{\rm v, L}|}\,\sigma(R,z) \textrm{ ,}
\end{equation}
and $\delta_{\rm v, L}$ is the linear underdensity threshold. Following the spherical evolution, nonlinear voids have expanded by a factor of $\simeq 1.7$ with respect to their linear stage. Consequently, the nonlinear (subscript "NL") void abundance becomes:
\begin{equation}
    \frac{\mathrm{d}n}{\mathrm{d}\ln R}\biggr|_{\rm SvdW} = \frac{\mathrm{d}n_{\rm L}}{\mathrm{d}\ln R_{\rm L}} \biggl |_{R_{\rm L} = R/1.7} \textrm{ .}
\end{equation}
The latter has been proposed by \citet{SvdW04} (SvdW, hereafter) and takes into account the suppression impact of the halo formation on the evolving population of voids (the \textit{void-in-cloud} phenomenon) expressing $\mathcal{D}$ as a function of the two threshold parameters $\delta_{\rm v}$ and $\delta_{\rm c}$. This translates into a shift of the linear model of Eq. \eqref{VSF_L} towards larger radii, without any change in amplitude, which means that in the SvdW model the comoving number density of voids is conserved in the nonlinear regime \citep{SvdW04}.

\citet{Jennings} pointed out that the latter is an incorrect assumption especially for large voids. Following this model, the cumulative fraction of volume occupied by voids exceeds the total volume of the Universe. Indeed, the SvdW model overpredicts the abundance of voids at any radius \citep{Jennings, RM17}, unless the threshold $\delta_{\rm v, L}$ is considered as a free parameter, which can be fine-tuned for different redshift and cosmic tracers by means of cosmological simulations. However, this severally affects the possibility of using the VSF as a cosmological probe.

In order to account for this problem, \citet{Jennings} proposed the volume-conserving size function model, or Vdn, in which the volume fraction occupied by cosmic voids is conserved in the transition from the linear to the nonlinear regime. This VSF model, which has been tested recently in many works \citep{Jennings, RM17, Ronconi19, Verza19, Contarini2019, Contarini2020,Contarini22Euc}, is defined as:
\begin{equation}\label{VdnModel}
\frac{\mathrm{d}n}{\mathrm{d}\ln R}\biggr|_{\rm Vdn} = \frac{\mathrm{d}n}{\mathrm{d}\ln R} \Biggr|_{\rm L} \frac{V(r_{\rm L})}{V(R)}\frac{\mathrm{d}\ln R_{\rm L}}{\mathrm{d}\ln R} \textrm{ .}
\end{equation}
In the VSF model, voids are defined as non-overlapping spheres embedding an internal underdensity contrast $\delta_{\rm v, NL}^{\rm DM}$ relative to the unbiased tracer distribution (DM particles). 

\begin{figure*}
	\includegraphics[width=\linewidth]{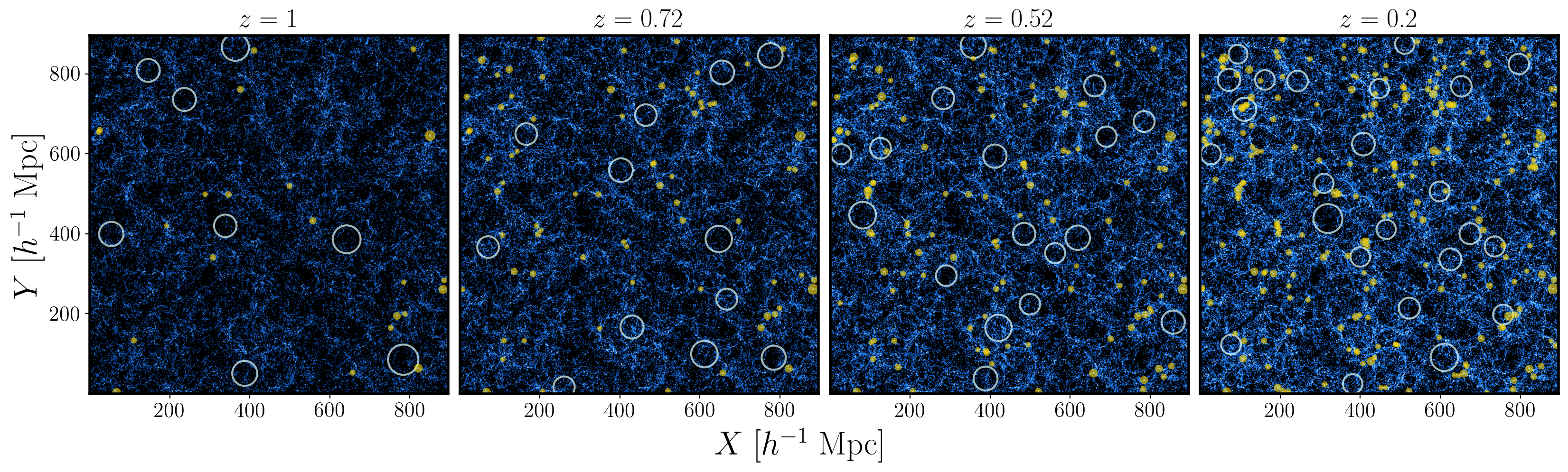}
	\centering
    \caption{The spatial distribution of simulated galaxies (blue dots), galaxy clusters (yellow dots) and voids (white circles) in the snapshots considered in this work, extracted from the \textit{Magneticum} simulations. The points representing galaxy clusters have dimensions proportional to their mass, while more massive galaxies have lighter colours. Void circles have have a radius corresponding to their size. The different panels refer to redshifts z=1, 0.72, 0.52, 0.2, from left to right. Here the cosmic objects considered follow the selections adopted in this work, i.e. $M_{*} \geq 10^{10}\ h^{-1} \ \rm{M}_\odot$ for galaxies, $M_{500c} \geq 10^{14}\ h^{-1} \rm{M}_\odot$ for clusters and $R_{\rm eff} \geq 3.6\ \lambda_{\rm mgs}(z)$ for voids. The axes are expressed in comoving coordinates,  considering a $60$ $h^{-1}\ \rm Mpc$ thick slice in the third coordinate.}
    \label{LSS}
\end{figure*}

In this work we consider voids identified in the distribution of simulated galaxies, which are biased tracers of the matter density field. In this case, the radii of the spherical voids predicted by the Vdn model have to be re-scaled in order to match the radii of voids identified in the nonlinear biased tracer distribution \citep{Jennings, Ronconi19, Contarini2019}. In our analysis, this is achieved by applying the cleaning algorithm developed by \citet{RM17} (see Section \ref{sec:dataprep}, for further details). The matching of the radii is obtained by re-scaling the model underdensity threshold $\delta_{\rm v, NL}^{\rm DM}$ by means of a bias factor $b$:
\begin{equation}
    \delta_{\rm v, NL}^{\rm tr} = b \delta_{\rm v, NL}^{\rm DM} \rm ,
\end{equation}
where $\delta_{\rm v, NL}^{\rm tr}$ is the tracer density contrast imposed to be embedded by voids in the biased field and selected during the void cleaning procedure. This threshold can be chosen from a wide range of negative values \citep[see][for further details]{Contarini2019}, as long as the corresponding linear density contrast is inserted in the Vdn model, as we will describe shortly. Following the studies of \citet{pollina2017,pollina2019}, \citet{Contarini2019, Contarini2020} showed that the bias $b$ does not coincide with the one computed on large scales, $b_\text{eff}$, measurable e.g. from the two-point correlation function (2PCF) of the tracer density field, demonstrating that the bias measured inside cosmic voids is linearly related to $b_\text{eff}$:
\begin{equation}\label{eq: F_beff}
    \mathcal{F}(b_{\text{eff}}) = B_\text{slope} b_\text{eff} + B_\text{off} \textrm{ ,}
\end{equation}
with $B_\text{slope}$ and $B_\text{off}$ being the slope and the offset of this relation, respectively. Consequently, the model nonlinear threshold $\delta_{\rm v, DM}^{\rm NL}$ can be computed as:
\begin{equation}
    \delta_{\rm v, NL}^{\rm DM} = \frac{\delta_{\rm v, NL}^{\rm tr}}{\mathcal{F}(b_{\text{eff}})} \textrm{ .}
\end{equation}
Moreover, \citet{Contarini2020} found a negligible dependence of this calibration relation on the considered cosmological model, that is a fundamental prerequisite to exploit the VSF as a cosmological probe.

Finally, the theoretical VSF model requires the corresponding linear underdensity threshold. Having $\delta_{\rm v, NL}^{\rm DM}$, one can convert it into its linear value via the following relation \citep{Bernardeau94}:
\begin{equation}
    \delta_{\rm v, L}^{\rm DM} = \mathcal{C}[1 - (1+\delta_{\rm v, NL}^{\rm DM})^{-1/\mathcal{C}}] \text{ ,}
\end{equation}
with $\mathcal{C} = 1.594$, the latter being exact for cosmologies with $\Lambda=0$, and very precise also for other values of $\Lambda$, especially when applied to the negative density contrasts.

\section{The Magneticum simulations}\label{MAGNETICUM}

In this work we make use of simulated galaxy and galaxy cluster catalogues extracted from the \textit{Magneticum Pathfinder}\footnote{\href{http://www.magneticum.org/}{http://www.magneticum.org/}} (Dolag et al., in preparation), a large set of cosmological, hydrodynamical simulations, having box volumes going from $(12\ h^{-1} \ \text{Mpc})^3$ (Box6) to $(2688\ h^{-1} \ \text{Mpc})^3$ (Box0). These boxes have different resolutions, and thus allow us to study the formation and evolution of both large-scale structures as well as phenomena that occur on smaller scales, from the motion of galaxies to the physics of gas inside them. The \textit{Magneticum} simulations have been run with the parallel code P-GADGET3, which is an updated version of the TreeSPH GADGET-2 code presented in \citet{Springel2005}. In this code, the gravitational forces are computed through a TreePM algorithm, while the hydrodynamics is modelled through an updated SPH algorithm \citep{Springel02,2016MNRAS.455.2110B}, from which it is possible to properly track the gas turbulence \citep{Dolag05}.

Furthermore the main baryonic physics phenomena are implemented in these hydrodynamic simulations, following the methods presented in \citet{Springel2003}, such as the cooling of gas, the star formation and supernovae feedback. In addition, black hole and active galactic nuclei feedbacks are included \citep{BHAGN, Fabjan10,2014MNRAS.442.2304H}, as well as thermal conduction \citep{2004ApJ...606L..97D}, stellar population and chemical enrichment models \citep{Tornatore2003, 2007MNRAS.382.1050T}. Following the latter, metals are produced by SNII, SNIa, and by low and intermediate mass stars in the asymptotic giant branch. The simulation assumes a Chabrier initial stellar mass function \citep{Chabrier03} and takes into account also the different release amount of metals and energy depending on the stellar mass \citep{Padovani93}. A more detailed description can be found in \citet{2014MNRAS.442.2304H} and \citet{2015ApJ...812...29T}.

The \textit{Magneticum} simulations have been carried out with the seven-year Wilkinson Microwave Anistropy Probe (WMAP7) cosmology \citep{WMAP7}, which we will consider as fiducial. In this scenario the main cosmological parameters are set to $\Omega_\mathrm{m} = 0.272$, $\Omega_\Lambda = 0.728$, $\sigma_8 = 0.809$, $H_0 = 70.4\ \text{km}\ \text{s}^{-1} \text{Mpc}^{-1}$, and $n_\mathrm{s} = 0.963$.
For our analysis we exploit the Box1a \citep[for details see][]{2016MNRAS.463.1797D}, which is a large simulation that follows the evolution of $2 \cdot 1526^3$ particles in a comoving volume of $(896\ h^{-1} \ \text{Mpc})^3$, with which it is possible to make a detailed statistical analysis of galaxy clusters and cosmic voids. This snapshot is built considering a mass resolution of DM and gas particles of $1.3 \times 10^{10}\ h^{-1} M_\odot$ and $2.6 \times 10^{9}\ h^{-1} M_\odot$, respectively. The simulation outputs, from which redshift-space cluster and void catalogues have been constructed, are selected at four redshifts, $z = 0.2, 0.52, 0.72, 1$. 

\subsection{Data preparation}\label{sec:dataprep}
We construct redshift-space mock catalogues of galaxies and galaxy clusters starting from their positions in the real space by following the same methodology used in \citet{Bianchi12} and \citet{Marulli11, Marulli12a, Marulli12b, MarulliVero}, which we briefly outline in the following. We set a virtual observer at redshift $z = 0$, and place the centre of the different simulation snapshots at a comoving distance $D_{\rm c}$, corresponding to their redshift:
\begin{equation}
    D_{\rm c}(z) = c \int_0^z \frac{\mathrm{d}z'_{\rm c}}{H(z'_{\rm c})} \text{ ,}
\end{equation}
where $c$ is the speed of light, $z_{\rm c}$ is the cosmological redshift due to the Hubble recession velocity and $H(z)$ is the Hubble parameter. The latter, for a $\Lambda$CDM cosmology and considering the contribution of radiation as negligible, becomes:
\begin{equation}
    H(z) = H_0 \bigl[\Omega_{\rm m} (1+z)^3 + \Omega_\Lambda\bigr]^{1/2} \text{ .}
\end{equation}

We then transform the comoving coordinates of real-space mock sources into angular positions and observed redshifts. For each galaxy in a given catalogue, the latter is computed from $z_{\rm c}$ as:
\begin{equation}
    z_{\rm obs} = z_{\rm c} + \frac{v_\parallel}{c}(1+z_{\rm c}) \text{ ,}
\end{equation}
with $v_{\parallel}$ being the galaxy centre of mass velocity projected along the line-of-sight. As in \citet{Marulli12b, MarulliVero}, we do not take into account any systematic error in the observed redshift, neither we consider geometrical distortions during the redshift-space mock catalogues generation, i.e. during this procedure we assume the correct, true underlying cosmology of the simulations.

In the \textit{Magneticum} simulations, haloes are identified with a Friend-of-Friend (FoF) algorithm \citep{Davis85}, with a linking length $b=0.16$ applied to the distribution of DM particles \citep{Dolag2009} while galaxy clusters are detected, starting from halo centres, as spherical overdensities. In our case, we consider spheres containing a linear overdensity $\Delta = 500\rho_c$, where $\rho_{\rm c} \equiv 3H^2 / 8\pi G$ is the critical density of the Universe at the considered redshift. We then select only clusters having mass $M_\text{500c} \geq 10^{14}\ h^{-1}\rm{M}_\odot$, choosing this mass selection in order to mimic the range of masses for galaxy clusters in real current and future surveys, like the extended Roentgen Survey with an Imaging Telescope Array (eROSITA) \citep{Merloni2012}. Also, for smaller masses we verified the presence of incompleteness in the cluster counts. We do not focus on any observable-mass relation in this work, and we consider cluster masses as a direct observable.

We underline that the adopted halo mass definition is not unique and that it is possible to switch to a different one under the assumption of a mean halo profile. For example, the mass cut off $M_\text{500c} \geq 10^{14}\ h^{-1}\ M_\odot$ would approximately translate into $M_\text{200c} \geq 1.4 \times 10^{14}\ h^{-1}\ M_\odot$ by considering a Navarro-Frenk-White profile with a concentration $\rm c = 4$  \citep[for further details see e.g.][]{Despali}. We therefore do not expect the halo mass definition to have a significant impact on the results presented in our analysis.

Voids are instead detected in the redshift-space distribution of galaxies. In the \textit{Magneticum} simulations galaxies are identified using a modified version of \textsc{subfind}, an algorithm aimed at extracting bound structures from a large $N$-body simulation characterised by the presence of gas and star particles \citep{Springel2001, Dolag2009}. We apply then a galaxy mass selection of $M_{*} \geq 10^{10}\ h^{-1} \ \rm{M}_\odot$ following the choice of \citet{MarulliVero}, who analysed the same cosmological simulations. We underline that there is not a direct link or correspondence between the galaxy cluster and galaxy mass cuts examined in this work. 

In each simulated box, redshift-space voids are identified with the public Void IDentification and Examination toolkit\footnote{\url{https://bitbucket.org/cosmicvoids/vide_public}}  \citep[{\small VIDE},][]{Sutter15}, which is based on the parameter-free void finding algorithm ZOnes Bordering On Voidness code \citep[{\small ZOBOV},][]{Neyrinck08}. The latter finds voids in a 3D distribution of tracer particles, without introducing any free parameter or assumptions about the void shape. 

The extracted void catalogues were then cleaned with the algorithm implemented inside the {\small CosmoBolognaLib}\footnote{\url{https://gitlab.com/federicomarulli/CosmoBolognaLib}} C++/Python libraries \citep{Marulli2016} and exploited in recent works \citep{RM17, Ronconi19, Contarini2019, Contarini2020, Contarini22Euc, Contarini22b, Contarini22c}. As already mentioned in Section \ref{VSFmodel}, the cleaning procedure is necessary because the VSF models described above consider voids as underdense, spherical, non-overlapping regions with a specific internal density contrast. Hence the aim of adopting this procedure is to prepare catalogues matching the theoretical definition used to develop the VSF model. The cleaning algorithm can be briefly described in three main steps: it removes non-relevant objects, i.e. voids with a radius outside of a given range and/or with a central contrast that is too high; it rescales the effective void radii, $R_{\rm eff}$, in order to make the voids embedding a specific density contrast; finally it checks for overlapping voids, eventually rejecting the ones with the highest central density. The impact of the cleaning procedure on the void abundances has been extensively tested in \citet{Ronconi19}, which applied it to a set of cosmological simulations at different redshifts having different spatial resolutions and box side lengths. 

Once the cleaning algorithm is applied, we select only voids having effective radii $R_{\rm eff} \geq 3.6\ \lambda_{\rm mgs}(z)$, with $\lambda_{\rm mgs}(z)$ being the mean galaxy separation of the catalogues from which voids are detected. This quantity corresponds to $\lambda_{\rm mgs} \simeq 6.6\ h^{-1}\rm Mpc$ at $z=0.2$ and increases up to $\lambda_{\rm mgs} \simeq 8.9\ h^{-1}\rm Mpc$ at $z=1$ because of the lower number of galaxies at higher redshifts.
We consider this extremely conservative radius cut in order to ensure reliable constraints on cosmological parameters (see Section \ref{test_constraints}). Indeed, with this selection we avoid considering the spatial scales characterised by incompleteness, i.e. those voids affected by the limited resolution of the simulation. Indeed, we verified that a less stringent radius selection would lead to cosmological test constraints slightly in disagreement with the true values of the \textit{Magneticum} simulation parameters. We refer the reader to previous works \citep[i.e.][]{Jennings, Ronconi19, Contarini2019} for further details on the impact of the spatial resolution on void number counts.

Figure \ref{LSS} shows the evolution of the large-scale distribution of galaxies (blue dots), galaxy clusters (yellow dots) and voids (white circles) inside slices of thickness 60 $h^{-1}\rm Mpc$, extracted from the \textit{Magneticum} snapshots. We also report simulated voids following the selection procedure described above, i.e. by applying cleaning and selection procedures. As one can note from this figure, the large-scale structure clustering increases with cosmic time, giving rise to more massive collapsed objects (i.e. galaxies and galaxy clusters) and, at the same time, to larger cosmic voids. In this representation however, some voids seem to be located in regions of galaxy crowding. This is just a projection effect given by the finite thickness of the shown slices. Moreover, some underdensity regions do not have an associated void: this can be explained in terms of the applied void selection, for which we considered only very large voids (see Section \ref{sec:dataprep}).

\section{Methods}\label{methods}
\subsection{From real-space to redshift-space}\label{sec:obs_eff}
The theoretical framework outlined in Section \ref{models} is only suitable for a real-space analysis. Considering redshift-space galaxy and galaxy cluster catalogues, it is necessary to correct for geometric and dynamic distortions. The former arise when the fiducial cosmology assumed in the analysis is different from the true one, while the latter are due to the peculiar motion of galaxies with respect to the Hubble flow. We will start our discussion by providing a model correction for the geometric distortions, which must be taken into account during the Bayesian Markov Chain Monte Carlo (MCMC) analysis. In fact, as the cosmological parameters change at each step of the MCMC, the mapping between redshift and comoving distance will be different, as well as the simulated box volume. The former is known as the Alcock--Paczynski effect  \citep[AP,][]{AP79} and influences the observed shape of large-scale structures, making them appearing elongated along the line of sight. Instead, the variation of the simulated volume affects the predicted number density of both clusters and voids.

Regarding clusters, we will include the effect of the varying simulated volume only, since the AP effect has a negligible impact on their masses. Viceversa, voids must be considered as spatially extended objects, hence the geometrical distortions must be taken into account when modelling their observed sizes.

Voids identified assuming the true cosmological model can be modelled as spherical objects having radius a $R$. However, it can be shown that assuming a fiducial cosmology different from the true one\footnote{Measurements computed in the fiducial cosmology are indicated here with primed symbols.}, the void radius $R'$ can be written as \citep[see][and references therein]{Contarini22Euc}:
\begin{equation}\label{AP_eff}
    R = q_{\parallel}^{1/3}q_{\bot}^{1/3} R'\rm ,
\end{equation}
where $q_{\parallel}$ and $q_{\bot}$ are defined via the following relations:
\begin{equation}\label{RSDs}
\begin{aligned}
    r_\parallel' &= \frac{H(z)}{H'(z)} r_\parallel = q_\parallel^{-1} r_\parallel \rm ,\\
    r_\bot' &= \frac{D_{\rm A}(z)}{D_{\rm A}'(z)} r_\bot = q_\bot^{-1} r_\bot \rm ,
\end{aligned}
\end{equation}
i.e. as the ratio between the Hubble parameter, $H(z)$, and the ratio between the comoving angular-distances, $D_{\rm A}(z)$, computed assuming the two cosmologies, respectively. In Eq. \eqref{RSDs}, $r_\parallel$ and $r_\bot$ are the parallel and perpendicular projections, with respect to the line of sight, of the distance vector connecting two points at redshift $z$. In applying this correction, we consider the mean redshift of the centers of voids included in each redshift interval. This approximation is accurate enough since the selected samples of voids cover small redshift ranges. %considered era usato 3 volte

As already mentioned, the other effect concerns the variation of the simulated box volume. This can be modelled as:
\begin{equation}\label{Vbox}
    V_{\rm box}(z) = L_{\rm box}^2 \cdot \Delta D_{\rm C}(z) \rm ,
\end{equation}
where $\Delta D_{\rm C}(z) = D_{\rm C}(z_{\rm max}) - D_{\rm C}(z_{\rm min})$ is the difference between the comoving distance of the furthest object and the nearest object in the simulated box. Here the redshift value depends on the simulated snapshot being considered.

With these corrections, the number counts of clusters in a given mass bin $\Delta M \equiv M_2 - M_1$ can be obtained as:
\begin{equation}
    N_{\rm cl}(\Delta M ; V_{\rm box}) = V_{\rm box} \int_{M_1}^{M_2} \text{d} M \frac{\mathrm{d}n}{\mathrm{d}M} \rm ,
\end{equation}
where $\mathrm{d}n / \mathrm{d}M$ is the HMF given by Eq. \eqref{HMFmodel}, and $V_{\rm box}$ is computed as in Eq. \eqref{Vbox}. Accordingly, the number counts of voids are computed from the VSF model of Eq. \eqref{VdnModel} as:
\begin{equation}
    N_{\rm v}(\Delta R ; V_{\rm box}) = V_{\rm box} \int_{R_1}^{R_2} \frac{\text{d} R}{R} \frac{\mathrm{d}n}{\mathrm{d}\ln R}\biggr|_{\rm Vdn} \rm ,
\end{equation}
where $\Delta R \equiv R_2 - R_1$ is a given void radius bin, with $R$ computed from Eq. \eqref{AP_eff}, i.e. taking into account the AP effect.

The inclusion of the geometric distortions in the HMF and VSF models, via the simulated volume change and the AP effect (for cosmic voids only), introduces a further dependence on the cosmological model. In particular, the total matter density parameter $\Omega_\mathrm{m}$ enters in the computation of the Hubble parameter $H(z)$ and in the comoving angular-distance $D_\mathrm{A}$ (see Eq. \ref{RSDs}), as well as in the comoving distance $D_\mathrm{C}$ (see Eq. \ref{Vbox}).

Finally, regarding dynamic distortions, we do not consider their impact on the HMF since the effect on the observed cluster masses can be considered negligible. Instead, it is fundamental to include the redshift-space distortions in the VSF model because the observed void radii undergoes a significant enlargement due to the coherent
outflow of tracers from the void centres \citep{Pisani2015b}. This effect can be encapsulated in the nuisance parameters, $B_\mathrm{slope}$ and $B_\mathrm{off}$, of the extended Vdn theory \citep{Contarini22Euc}. In our analysis, this will be performed in Section \ref{sec:calibrationHMF}, during the calibration of the VSF with the measured redshift-space void number counts.

\subsection{Bayesian analysis}\label{sec:bayes}
Having extracted our data sets $\mathcal{D}$, consisting of cluster and void counts measured from the simulations, we follow a Bayesian approach to sample the posterior distribution of two fundamental cosmological parameters, i.e. the total matter density parameter, $\Omega_{\rm m}$, and the $z = 0$ amplitude of the matter power spectrum, $\sigma_8$. Let $\Theta = (\Omega_{\rm m}, \sigma_8, \textbf{q})$ denote the parameter vector, with $\textbf{q}$ being the set of the internal HMF and VSF model parameters. Then the posterior distribution of $\Theta$, given a cosmological model, can be computed from the Bayes's theorem:
\begin{equation}\label{bayes_th}
    \mathcal{P}(\Theta | \mathcal{D}) \propto \mathcal{L}(\mathcal{D} | \Theta) p(\Theta)\rm ,
\end{equation}
where $\mathcal{L}(\mathcal{D} | \Theta)$ is the likelihood function and $p(\Theta)$ the priors considered for the parameters.

We then make use of MCMC methods for the posterior sampling, taking into account the correlation between different data sets where needed (see Section \ref{cross-cov}). In the general case, i.e. where the data set covariance matrix $\mathcal{C}$ is non-diagonal, the following Gaussian likelihood is considered:
\begin{equation}
    \label{multivariate_gaussian}
\mathcal{L}(\mathcal{D} | \Theta) = \prod_{i,j} \frac{1}{(2\pi)^{n/2}\sqrt{|\mathcal{C}|}}\exp{\Bigl(-\frac{1}{2}\chi_{ij}^2\Bigr)} \textrm{ ,}
\end{equation}
where 
\begin{equation}\label{eq:chi2}
\chi_{ij}^2 = (N_{ij} - \mu_{ij})^\text{T} \mathcal{C}^{-1} (N_{ij} - \mu_{ij}) \textrm{ .}
\end{equation}
Here $n$ is the number of items in the data set, $N_{ij} \equiv
N(\xi_{i}, z_j | \mathcal{D})$ is the number counts for the given
observable $\xi$ in the $i$-th bin (mass or radius) at redshift $z_j$,
and $\mu_{ij} \equiv N(\xi_i, z_j | \Theta)$ is the expectation value
in the cosmological model considered. Note that the covariance matrix of a given probe (i.e. HMF or VSF, in our case), $\mathcal{C} \equiv \mathcal{C}(\xi_i, z_j)$, depends both on the measured statistic in the given bin, $\xi_i$, and on the mean redshift of the bin, $z_j$. In this work we exploit the combination between different posterior distributions, obtained separately from our cosmological probes. This can be achieved via the product of their likelihood functions, or more generally their posterior product.

\subsection{Correlations between cluster and void counts}\label{cross-cov}
To check for possible correlations between cluster and void counts, we compute the cross-covariance matrix of the two probes considered, employing the jackknife sampling technique \citep{Parr80, Efron82, Escoffier16}. The latter is used to generate new mock samples that are representative of an underlying population, i.e. the analysed catalogues in this work, whose statistics we want to estimate. This technique works by sequentially removing one sub-sample from the entire data set, computing the desired statistics on the remaining samples.

\begin{figure}
	\includegraphics[width=\columnwidth]{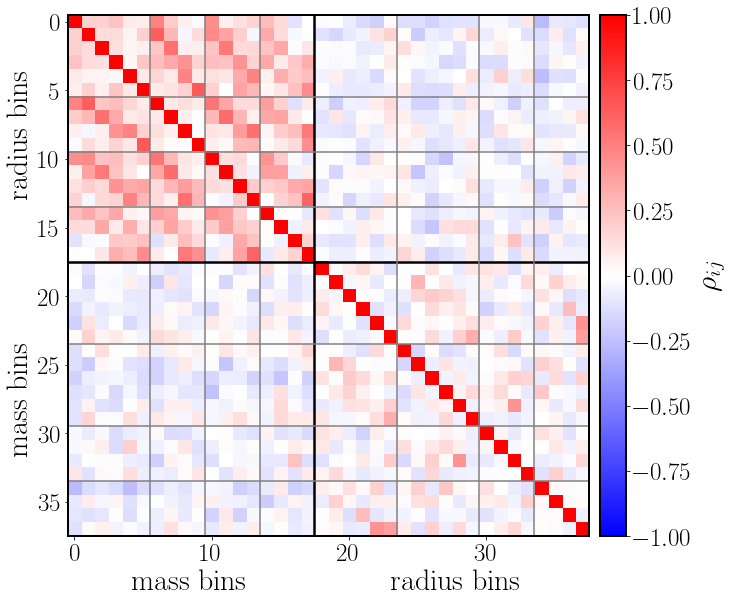}
    \caption{Jackknife cross-correlation
      matrix for galaxy cluster and void counts, for the mass and
      radius selections considered in the paper. The colorbar on the right represents the Pearson correlation coefficient associated to cluster mass and void radius bins. Data sets at different redshifts are separated by grey solid lines, while mass bins and radius bins are divided by solid black lines.}
    \label{cross_corr}
\end{figure}

For each redshift considered, we divide the cluster and void catalogues into $N_{\rm sub} = 125$ equally sized sub-regions. A value for the number counts of galaxy clusters $N_\alpha$ and cosmic voids $N_\beta$ can be associated with $\alpha \sim (i_M , i_z)$, a generic bin in mass and redshift, and $\beta \sim (i_R, i_z)$, a generic bin in radius and redshift. The jackknife covariance element between two generic bins $\alpha$ and $\beta$ can then be estimated as \citep[see e.g.][]{Efron82}:
\begin{equation}\label{jk_crosscov}
    \mathcal{C}^{\text{jk}}_{\alpha\beta} = \frac{N_{\text{sub}}}{N_{\text{sub}} - 1} \sum_{i=1}^{N_{\text{sub}}} \bigl(N^{\text{jk}}_\alpha (i) - \bar{N}^{\text{jk}}_\alpha\bigr)\bigl(N^{\text{jk}}_\beta(i) - \bar{N}^{\text{jk}}_\beta\bigr) \textrm{,}
\end{equation}
where:
\begin{equation}
    \bar{N}^{\text{jk}}_\alpha = \frac{1}{N_{\text{sub}}} \sum_{i=1}^{N_{\text{sub}}} N^{\text{jk}}_\alpha (i)
\end{equation}
is the average of the jackknife counts in a specific bin and $
N^{\text{jk}}_\alpha (i) = N^{\text{tot}}_\alpha - N_\alpha(i)$ is a jackknife sample, defined as the difference between the number counts in the total volume and the ones in the volume of one sub-sample. In this way, the cross-covariance matrix between the number counts of galaxy clusters and cosmic voids is a block matrix, and can be written as:

\begin{equation}
    \mathcal{C}^{\text{jk}} = \left(\begin{array}{c|c}
\mathcal{C}_{\rm HH} & \mathcal{C}_{\rm VH} \\
\hline
\mathcal{C}_{\rm HV} & \mathcal{C}_{\rm VV}
\end{array}\right) \textrm{ ,}
\end{equation}
where $\mathcal{C}_{\rm HH}, \mathcal{C}_{\rm VH}, \mathcal{C}_{\rm HV}$ and $\mathcal{C}_{\rm VV}$ are the covariance matrices between cluster mass bins, cluster mass and void radius bins (and vice-versa) and void radius bins at all redshifts ($z = 0.2, 0.52, 0.72, 1$), respectively. Since the covariance matrix has been computed using a limited set of mock catalogues, we include the statistical corrections suggested by \citet{Percival14} to improve the accuracy in the estimation of the model parameters. Additionally, we apply the prescriptions by \citet{Hartlap07} to correct for the numerical uncertainties in the inversion of the covariance matrix, needed in Eq. \eqref{eq:chi2}. We note some level of noise in the computed covariance matrix. We discuss the stability of our results against the variation of $N_{\rm sub}$ in Appendix \ref{appendix:a}.

In Figure \ref{cross_corr} we present the cross-correlation matrix between our two probes, computed for each bin as the Pearson correlation coefficients:
\begin{equation}
    \rho_{ij} = \frac{\mathcal{C}_{ij}}{\sqrt{\mathcal{C}_{ii} \times \mathcal{C}_{jj}}} \textrm{ .}
\end{equation}
We clearly note that the internal correlation matrix of galaxy cluster counts, i.e. the upper left block matrix in Figure \ref{cross_corr}
presents non-vanishing off-diagonal elements. These features are due to the fact that we are analysing the redshift evolution of the same large-scale structure, so we are considering the same objects evolving with time. We checked that these correlation features in fact disappear when estimating again the covariance matrix for cluster number counts, but resampling from different cosmological sub-boxes at each redshift. Despite this approach would be more accurate in reproducing the behaviour of real cosmic objects, it severely reduces the measured cluster number counts and so would negatively affects the statistical relevance of the analysis. For this reason we applied this methodology as a consistency test only. On the contrary, we do not get evident correlation features in the void count correlation matrix. We attribute this fact to both the noisier data and the effect of the cleaning procedure, which reduces the possibility of following the
evolution of the same void sample with cosmic time. Notably, we can appreciate how the level of correlation of the off-diagonal block matrices $\mathcal{C}_{\rm VH}$ and $\mathcal{C}_{\rm HV}$ is consistent with zero, confirming the high statistical independence of cluster and void number counts, as also found in previous works \citep{Sahlen16, Bayer21, Kreisch21}.

\section{Results}\label{results}
Before comparing the simulated data sets extracted from the \textit{Magneticum} simulations with the theoretical models described in Section \ref{models}, a self-calibration (i.e. on the same simulation outputs, considering the fiducial cosmology) of the latter is necessary, because of the differences in the halo definitions between the simulations employed here and those assumed by the theoretical HMF model, as already explained in Section \ref{sec:clustercounts}. In this case, the calibration will allow us to determine the posterior distribution for the coefficients that appears in Eq. \eqref{fitting}. Consequently, we will marginalise over the latter in order to find the posterior distributions for the cosmological parameters $\Omega_{\rm m}$ and $\sigma_8$ (see Section \ref{test_constraints}).

\begin{figure}
	\includegraphics[width=\columnwidth]{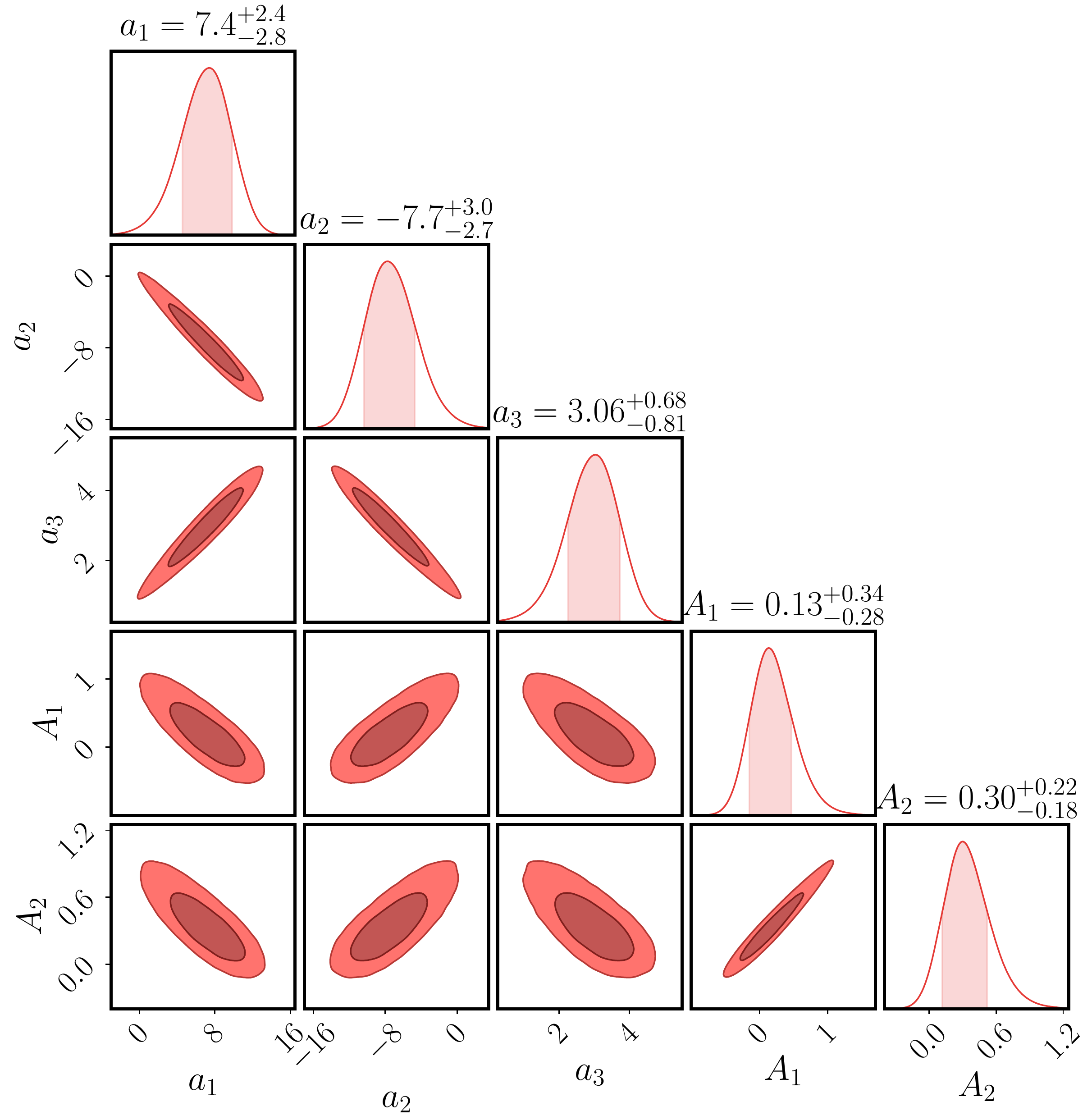}
    \caption{Posterior probability distribution of the coefficients of the fitting relation parameters ($a, A$) reported in Eq. \eqref{fitting}, for a cluster mass selection of $M_{500c} \geq 10^{14}\ h^{-1} \ \rm{M}_\odot$. Dark and light areas show the $68\%$ and $95\%$ confidence regions, respectively. On the top of each columns we report the projected 1D marginal posterior distributions, with the associated maximum posterior values and relative $1\sigma$ uncertainties.}
    \label{calibrationHMF}
\end{figure}

\begin{figure}
	\includegraphics[width=0.9\columnwidth]{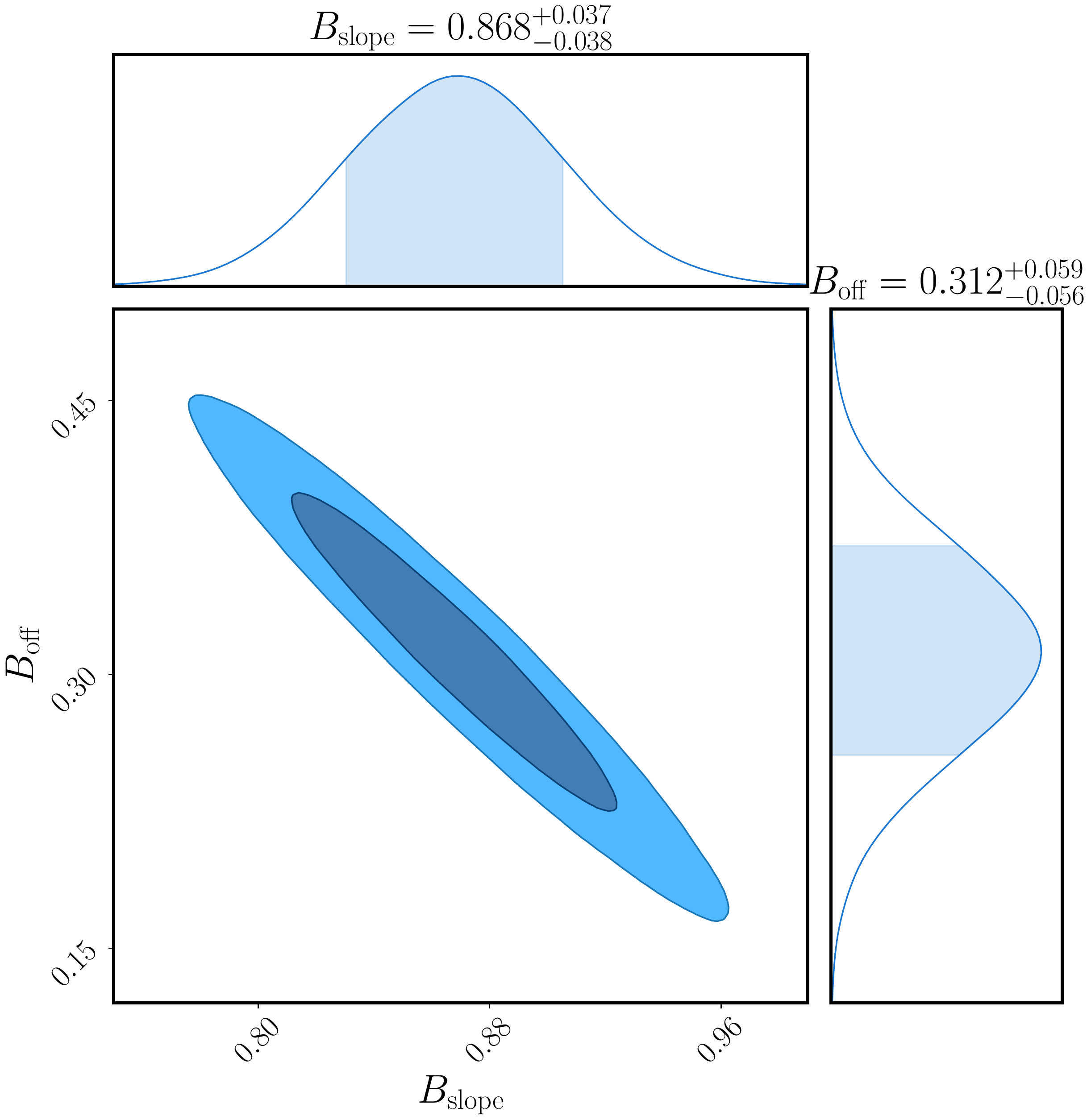}
    \caption{Posterior probability distribution of $B_\text{slope}$ and $B_\text{off}$, coefficients of the $\mathcal{F}(b_\text{eff})$ relation, reported in Eq. \eqref{eq: F_beff}. Dark and light areas show
the 68\% and 95\% confidence regions, while at the top of each column we report the projected 1D distributions, with the values associated to the maximum of the posterior and the relative $1\sigma$ uncertainties.}
    \label{VSF_calib}
\end{figure}

Moreover, also the VSF model has to be calibrated \citep{Contarini2019,Contarini2020,Contarini22Euc}. We make use of the Vdn model described in Section \ref{VSFmodel}, modifying properly the underdensity threshold that enters in the model, as reported in Eq. \eqref{eq: F_beff}. Specifically, the calibration we  perform on the VSF model is to assess the posterior distribution of the free parameters of the function $\mathcal{F}(b_\mathrm{eff})$, i.e. $B_\mathrm{slope}$ and $B_\mathrm{off}$.

For both the calibrations and constraints on cosmological parameters, we fit the joint data set at all the redshift considered with the model presented in Section \ref{models}. In particular, we compute the constraints by concatenating the data sets concerning the two probes analysed at different redshifts into a unique vector, constructing the likelihood with the correct covariance matrix for the given cosmological probe.

\subsection{Model self-calibrations}\label{sec:calibrationHMF}
The calibration performed in \citet{Despali} considered DM-only cosmological simulations, while the \textit{Magneticum} simulations also follow the baryonic component. Moreover, \citet{Despali} defined the galaxy clusters with a halo finder method different from the one used in this work. For these reasons, we have to re-calibrate the parameters of the HMF model, by assuming the functional relation given by Eq. \eqref{fitting}. This translates into finding other coefficients for the relations $a = a(x), p = p(x)$ and $A = A(x)$. We chose to consider $p$ as fixed to the value provided by \citet{Despali} in the most general fitting case, that is $p = 0.2536$. We made this choice because $p$ significantly affects only the low mass end of the HMF, and we cannot precisely constrain it due to the mass selection applied and the simulation mass resolution.

We expressed the general form of the fitting formula Eq. \eqref{fitting} as:
\begin{equation}\label{newfitting}
\begin{aligned}
a &= a_1 x^2 + a_2x + a_3\\
A &= -A_1 x + A_2\\
p &= 0.2536 \textrm{ ,}
\end{aligned}
\end{equation}
and we calibrate the new relations for the model parameters by performing a Bayesian MCMC analysis on the simulated HMF, considering flat priors for the five free parameters ($a_1, a_2, a_3, A_1, A_2$) and sampling their posterior distributions. Since the correlation matrix associated to the HMF (upper left block matrix in Figure \ref{cross_corr}) is non-diagonal, during the calibration procedure we consider a Gaussian likelihood as in Eq. \eqref{multivariate_gaussian}.

The result of the calibration is presented in Figure \ref{calibrationHMF}, which shows the posterior distribution of the coefficients of Eq. \eqref{newfitting}. We find strong correlations between the different calibration parameters. Moreover the marginalised 1D posterior distributions present an approximately Gaussian shape, with a low degree of skewness, the latter being more prominent for the $a_i$ parameters. While $A_1$ and $A_2$ result statistically in agreement with the values obtained from the calibration performed in \citet{Despali} (see Eq. \ref{fitting}), the values for the $a_i$ coefficients are in  $>2\sigma$ disagreement with respect to their recovered values. This is an expected outcome due to the different adopted halo definitions \citep[see][for further details]{Despali}, and also to the effect of the baryonic physics \citep[see e.g.][]{Bocquet16, Ragagnin21}, which is included in the simulations analysed in this work.

\begin{figure}
    \centering
	\includegraphics[width=1.1\columnwidth]{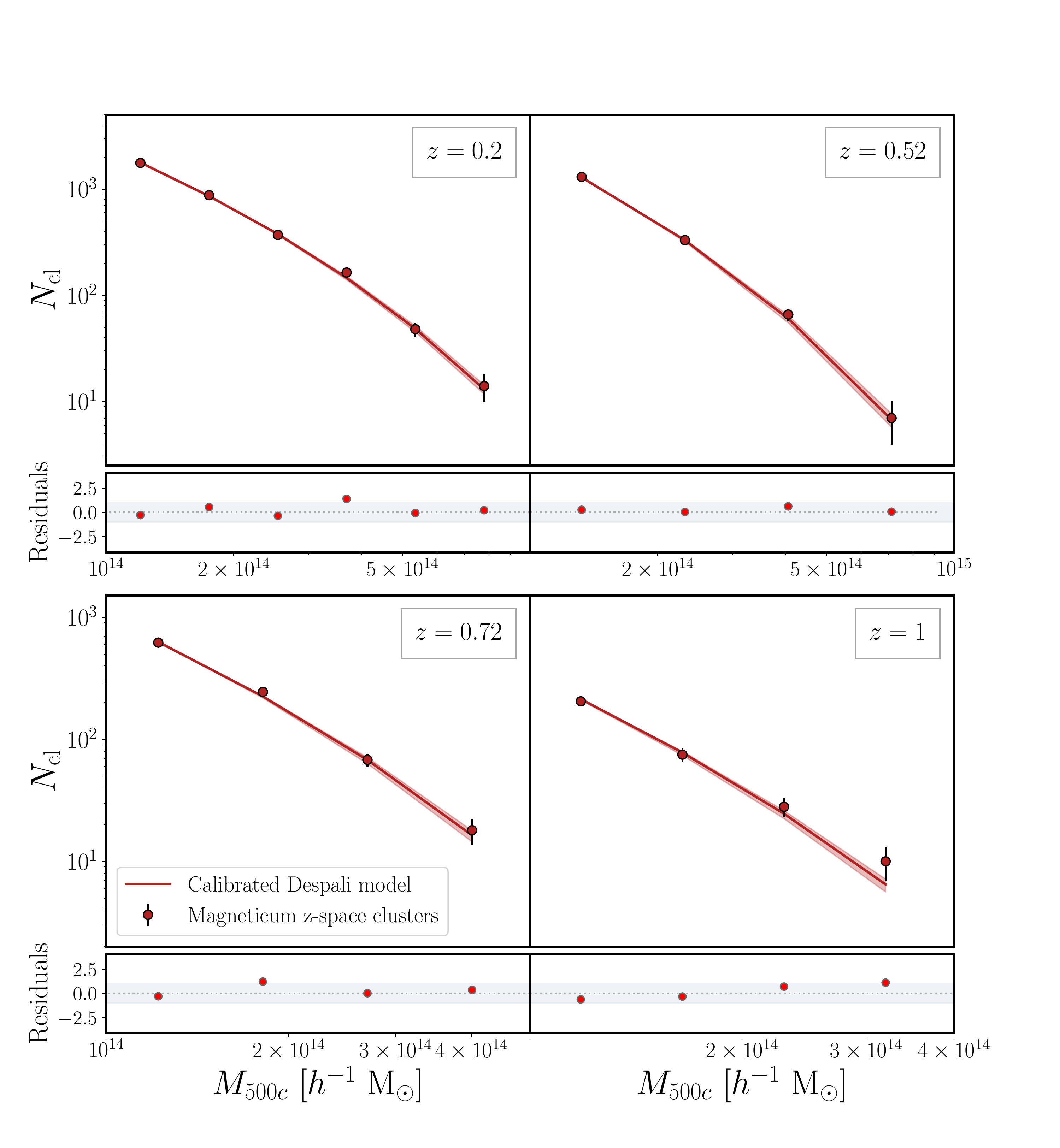}
    \caption{Measured number counts of galaxy clusters (red dots), from the \textit{Magneticum} (Box1a) at redshifts $z = 0.2, 0.52, 0.72, 1$, for ${M_{500\rm c} \geq 10^{14}\ h^{-1} \rm{M}_\odot}$. Red solid lines represents the HMF model from \citet{Despali} after the parameter re-calibration. Lower sub-panels report the residuals of the cluster counts, computed as the difference of the measured data and the HMF model, in units of the data errors. The light grey bands represent the $1\sigma$ intervals for the residuals.}
    \label{cluster_counts}
\end{figure}

\begin{figure}
    \centering
	\includegraphics[width=1.1\columnwidth]{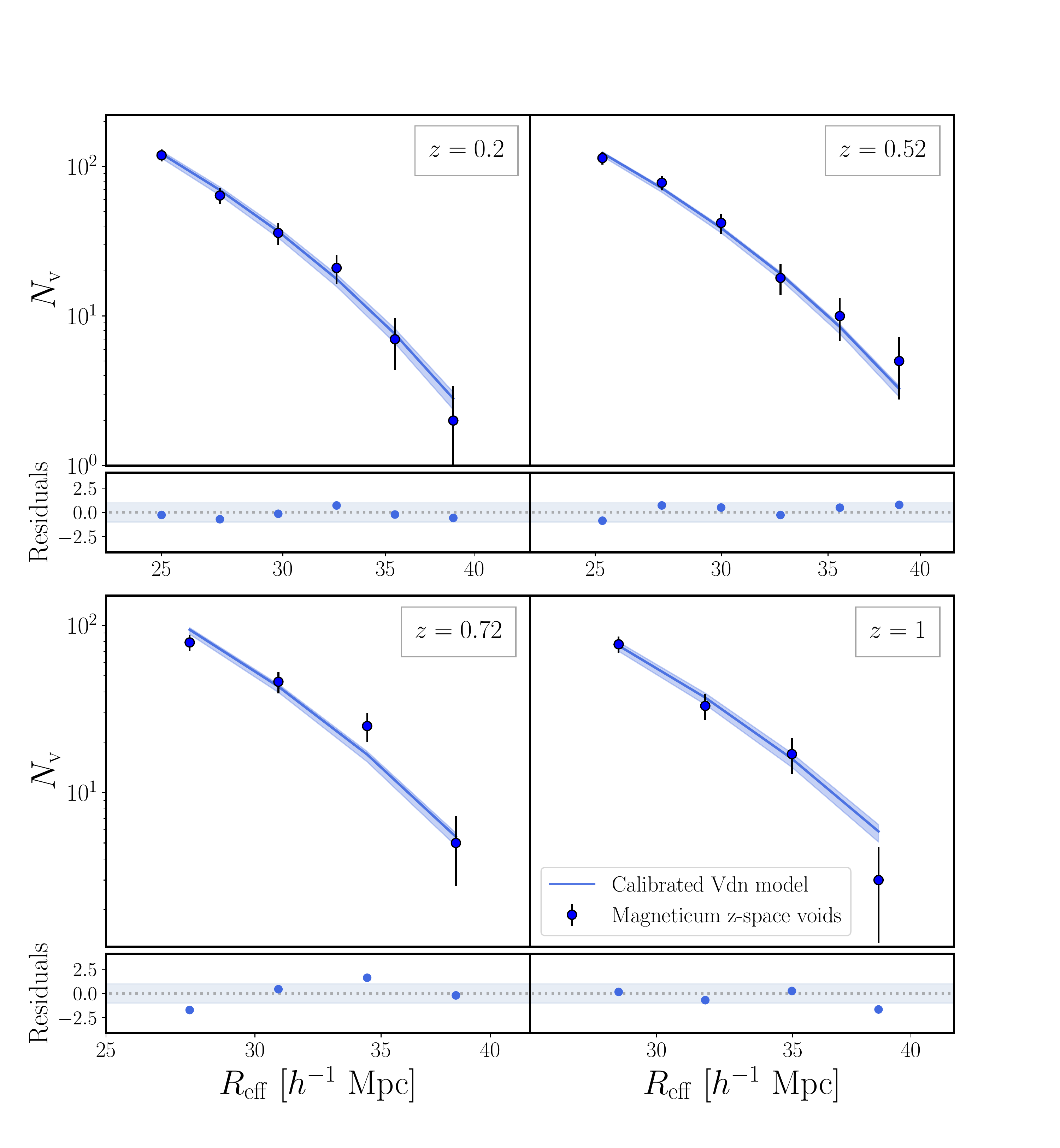}
    \caption{Measured number counts of voids (blue dots) identified in the distribution of galaxies, from the \textit{Magneticum} simulation (Box1a) at redshifts $z = 0.2, 0.52, 0.72, 1$. The blue lines represent the corresponding predictions of the Vdn model, extended and calibrated according to the prescriptions provided in Section \ref{VSFmodel}. The lower sub-panels report the residuals of the void counts, computed as the difference of the measured data and the VSF model, in units of the data errors. The light blue bands represent the $1\sigma$ intervals for the residuals.}
    \label{VSF_calib_counts}
\end{figure}

Similarly to what we have done for the HMF parameters, here the VSF model calibration consists in finding new values for the parameters $B_\text{slope}$ and $B_\text{off}$ of Eq. \eqref{eq: F_beff}, by sampling their posterior distributions via a MCMC analysis, with the cosmological parameters fixed to the simulation true values.

In order to do this, we first compute the effective large-scale bias, $b_\text{eff}$, following the method delined in \citet{Marulli13} and \citet{Marulli2018}. In particular, we construct a 10-time denser random catalogue with the same properties of the \textit{Magneticum} simulation boxes and measure the 2PCF of real-space galaxies making use of the Landy-Szalay estimator \citep{LandySzalay93}.  We then sample the posterior distribution of the free parameter of the model, $b_\text{eff}$, following the same method described in \citet[Appendix A]{Contarini2019}. We obtain $b_{\rm eff} = 1.321 \pm 0.007, 1.549 \pm 0.008, 1.714 \pm 0.009, 1.98 \pm 0.01$ for the four redshifts considered in our analysis, modelling the galaxy 2PCF in the range $[20-40]\ h^{-1}\rm Mpc$. Then we calibrate the VSF model by fitting the measured void number counts. In particular we consider flat priors for $B_{\rm slope}$ and $B_{\rm off}$, marginalising over the values of $b_{\rm eff}$ by assuming 1D Gaussian priors centred on the computed mean values and with standard deviation equal to their reported errors. As for cluster counts, we considered a Gaussian likelihood (Eq. \ref{multivariate_gaussian}) with the covariance matrix computed in Section \ref{cross-cov}. 

We present the result of this calibration in Figure \ref{VSF_calib}, where we show the posterior distribution of $B_{\rm slope}$ and $B_{\rm off}$, which have both approximately Gaussian shape. Also in this case we found a strong correlation between the parameters, with degeneracy direction consistent with recent works \citep{Contarini2020, Contarini22Euc}. We find a discrepancy with the results found in \citet{Contarini22Euc}, i.e. $B_{\rm slope} = 0.96 \pm 0.04$ and $B_{\rm off} = 0.44 \pm 0.07$. This is due to the different mass tracer type and selection used by these authors: in \citet{Contarini22Euc} voids are traced by H$\alpha$ galaxies, simulated by means of a halo occupation distribution algorithm and selected according to a magnitude cut \citep[see][for further details]{Contarini22Euc}. In this work we analyse instead voids identified in hydrodynamical mock samples of galaxies, selected with a threshold in stellar mass.

\subsection{Cosmological test constraints}\label{test_constraints}
We can now present the measurements of the number counts of both galaxy clusters and cosmic voids extracted from redshift-space catalogues, and compare these with the theoretical predictions given by the calibrated models. We rejected voids that are too close to the simulation boundaries, considering an edge of $15\ h^{-1}$ Mpc, since the cleaning algorithm cannot rescale them accurately. We present the measured number counts of galaxy clusters and cosmic voids in Figures \ref{cluster_counts} and \ref{VSF_calib_counts}, respectively, for the four redshift bins considered. The bars represent the Poissonian errors related to the data while the $\pm 1\sigma$ uncertainty of theoretical models are shown as shaded regions. From the residuals, computed as the difference between the measured data points and the theoretical models in units of the data error, one can appreciate how the measured number counts are accurately fitted by the theoretical models. Indeed, residuals are in all cases within $1\sigma$.

Once measured the HMF and VSF models from the simulated data sets, we extract test constraints on the cosmological parameters $\Omega_{\rm m}$ and $\sigma_8$. As already explained, we consider cluster and void abundances as a function of the mass and radius, respectively, as statistically independent, meaning that we do not take into account the cross-covariance matrix between the two probes. We perform a Bayesian MCMC analysis, considering flat priors on $\Omega_{\rm m}$ and $\sigma_8$ with ranges $[0.1-0.5]$ and $[0.5-1.0]$, respectively, and marginalising over the non-cosmological parameters, i.e. considering them as nuisances. 
Moreover, during the variation of $\Omega_{\rm m}$ in the MCMC, we imposed the flat spatial geometry of the Universe by rescaling the value of the CDM component as ${\Omega_{\rm cdm}=\Omega_{\rm m}-\Omega_{\rm b}}$ in order to hold true the condition $\Omega_{\rm m}+\Omega_{\Lambda}=1$.

The correlation between nuisance parameters of the models, i.e. $(a_i , A_i)$ for the HMF and $(B_{\rm slope} , B_{\rm off})$ for the VSF, is taken into account by considering a multivariate normal distribution as corresponding prior, centred on the parameter average values obtained from the calibrations (Sections \ref{sec:calibrationHMF}). As already mentioned, we use the jackknife covariance matrix for both cluster and void counts. Then, we progressively stack the cluster and void abundance data sets relative to different simulation snapshots, sampling the joint posterior distribution of $\Omega_{\rm m}$ and $\sigma_8$  for increasing redshift ranges. This procedure is aimed at assessing the constraining power on the growth of cosmic structures deriving from the synergy of cluster and voids number counts at different cosmic times.

\begin{figure}
	\includegraphics[width=1.1\columnwidth]{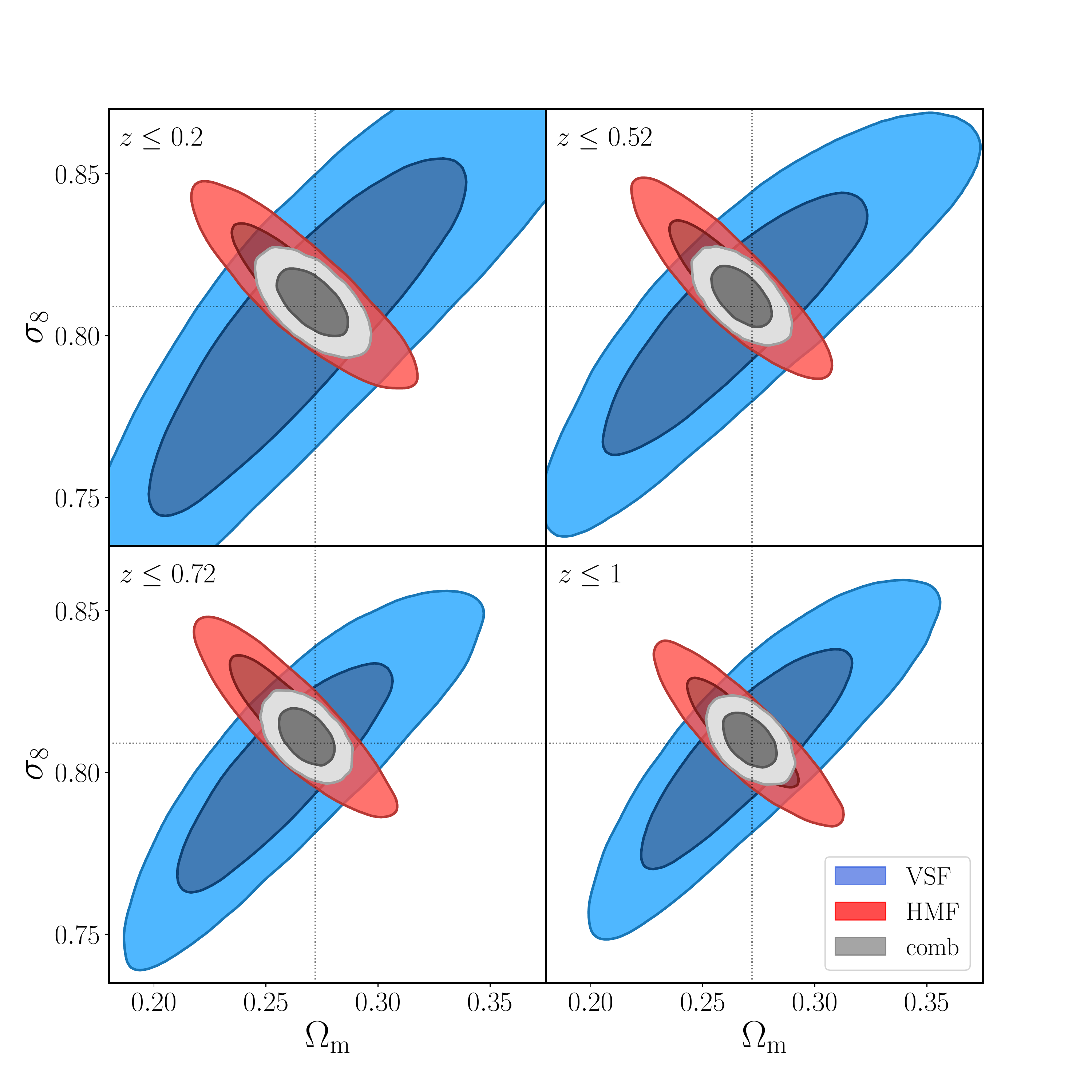}
    \caption{Posterior probability distribution on $\Omega_{\rm m}$ and $\sigma_8$ from VSF (blue), HMF (red) and their combination (gray), for clusters and voids extracted from \textit{Magneticum} simulations, considering their joint analysis up to a given maximum redshift, i.e. $z = 0.2, 0.52, 0.72, 1$. The dashed black lines show the truth values of the cosmological parameters of the \textit{Magneticum} simulations.}
    \label{combination}
\end{figure}

Figure \ref{combination} shows the resulting contours for the HMF model considering redshift-space clusters with masses $M_{500\rm c} \geq 10^{14}\ h^{-1} \rm{M}_\odot$, for the VSF model with voids having radius $R_{\rm eff} \geq 3.6\ \lambda_{\rm mgs}(z)$, and for their combination, considering a gradually increasing maximum redshift up to $z = 1$. In all cases, the HMF and VSF contours present different orientations in the $\Omega_{\rm m}$-$\sigma_8$ parameter space. In particular they are almost perpendicular, which is a powerful feature when we consider their combination. 
As expected, the precision of the constraints increases by adding the cosmological information of different Universe epochs. We underline indeed how the combined confidence contour shrinks by going from $z \leq 0.2$ to $z \leq 1$, remaining however perfectly in agreement with the true cosmological parameters of the simulations. We also notice that the VSF gains more constraining power with the extension of the redshift range with respect to the HMF. Thanks to the negligible correlation between the void samples at different redshifts, the combination of the corresponding VSF constraints results indeed very effective. For the HMF, instead, adding cosmological information from higher redshifts is less advantageous mainly because of the correlation between the galaxy clusters belonging to the different simulation snapshots: as already explained, this is taken into account by including in the modelling of cluster counts the full-covariance matrix (see Section \ref{cross-cov}) and results in a milder increase of the constraining power since it derives from the joint analysis of dependent data sets.

The final combined constraints on $\Omega_{\rm m}$ and $\sigma_8$, obtained by considering all the data sets up to $z = 1$, are the following:
\begin{equation}\label{final_constr}
\begin{aligned}
    \Omega_{\rm m} = 0.270 \pm 0.007\\
    \sigma_8 = 0.809 \pm 0.005 \rm ,
\end{aligned}
\end{equation}
which, as well as those coming from the single probes, are well centered on the truth values of the simulation, $\Omega_{\rm m} = 0.272$ and $\sigma_8 = 0.809$. We notice that, despite the maximum of the posterior distribution deriving from the VSF is slightly shifted towards lower values in the $\Omega_{\rm m}$-$\sigma_8$ space, the true simulation cosmological parameters are widely included in the the corresponding 68$\%$ confidence region.

To have an easier comparison between the combination constraints and the ones from HMF and VSF alone, we derive 1D constraints also for $S_8 \equiv \sigma_8 \sqrt{\Omega_{\rm m}/0.3}$, which are reported in Table \ref{tab:S8_constraints}. First of all we note how the truth value $S_8 = 0.77$ is well recovered within the 68$\%$ confidence region for all the redshift ranges considered in the table.
Then, in order to evaluate the contribution of the VSF in the performed probe combination, we show how much the precision on $S_8$ increases with respect to that computed from the HMF alone.
This percentage improvement, denoted here as $\mathcal{I}_{\rm HMF}$, is computed as the ratio between the relative errors on $S_8$ from HMF and from the combined analysis. We find an improvement on the constraining power of about $60\%$ compared to that obtained with the HMF alone, almost independent of the maximum redshift considered in the analysis. 

This result highlights the impressive contribution of the VSF in enhancing the constraining power derived with standard probes like the HMF, moreover it underlines the fundamental importance of exploiting complementary probes in future redshift surveys to achieve precise and independent cosmological constraints.

\begin{table}
    \renewcommand{\arraystretch}{1.5}
	\centering
	\caption{Test constraints up to a given redshift (first column) on the $S_8$ parameter. The reported errors represents the $68\%$ confidence regions on $S_8$. In the last column $\mathcal{I}_{\rm HMF}$ represents the improvement on the constraining power achieved by the combination of HMF and VSF compared to the HMF alone, computed as the ratio between the relative errors on $S_8$ from HMF and from HMF+VSF.}	
	\label{tab:S8_constraints}
	\begin{tabular}{lcccc} 
		\hline
		\hline
		Redshift & $S_8(\rm HMF)$ &  $S_8(\rm VSF)$ & $S_8(\rm HMF+VSF)$ & $\mathcal{I}_{\rm HMF}$ \\
		\hline
		\hline
		$z\leq 0.2$ & $0.763^{+0.019}_{-0.017}$ & $0.75^{+0.10}_{-0.09}$ & $0.77\pm 0.01$ & $67\%$ \\
		$z\leq 0.52$ &  $0.760^{+0.017}_{-0.014}$ & $0.75 \pm 0.07$ & $0.77 \pm 0.01$ & $65\%$ \\
		$z\leq 0.72$ & $0.760^{+0.016}_{-0.015}$ & $0.74 \pm 0.06$ & $0.767^{+0.009}_{-0.01}$ & $63\%$ \\
		$z\leq 1$ & $0.765^{+0.017}_{-0.013}$ & $0.76 \pm 0.06$ & $0.77 \pm 0.01$ & $67\%$ \\
		\hline
	\end{tabular}
\end{table}

\section{Conclusions}\label{conclusions}
In this work we evaluated the constraining power of the combination between two large-scale cosmological probes, namely the number counts of galaxy clusters and cosmic voids. We studied the potential constraints achievable by sampling the posterior distribution of the parameters $\Omega_{\rm m}$ and $\sigma_8$ through the analysis of the
redshift-space cluster and void catalogues extracted from the \textit{Magneticum} Box1a simulation, in four redshift snapshots ($0.2 \leq z \leq 1$). Since this simulation takes into account the baryonic physics, we re-calibrated the HMF and VSF parameter models to minimise systematic uncertainties. We considered only galaxy clusters having masses $M_{500c} \geq 10^{14}\ h^{-1} \rm{M}_\odot$, and voids with an effective radii $R_{\rm eff} \geq 3.6\ \lambda_{\rm mgs}(z)$, with $\lambda_{\rm mgs}(z)$ being the mean galaxy separation from which voids are detected. We chose these conservative selections to make our mock catalogues similar to those expected from large-scale surveys and, at the same time, to not be influenced by the resolution of the simulation. We computed the cross-covariance matrix between the two probes, finding no statistically relevant correlations between cluster
mass and void radius bins at different redshifts. By looking at the cross-covariance matrix we found, instead, correlation features within cluster mass bins at different redshifts.

The cross-correlation terms are dominated by noise, so we have decided not to include their contribution in the calculation of the final constraints. It would be interesting to repeat our analysis with a larger set of mock catalogues. This would allow us to better investigate the cross-correlation terms, thus possibly improve the accuracy of the results. Moreover, it would be interesting to investigate the impact of cross-correlations between different redshift bins. This is beyond the scope of the current analysis, and it will be investigated in a forthcoming work. 

The most remarkable result of our analysis concerns the high constraining power derived from the combination of the HMF and VSF on the $\Omega_{\rm m}$ - $\sigma_8$ space: being the confidence contours computed with these probes highly perpendicular and their cross-correlation negligible, the cluster and void number count joint analysis is of great effectiveness. In particular, we found that the constraining power given by the cluster counts improves by about $60\%$ when combined with void counts. This highlights the important contribution that void counts can provide to reduce the relative errors on the cosmological parameters related to the growth of large-scale structures, i.e. $\Omega_{\rm m}$, $\sigma_8$ and the derived quantity $S_8$. Notably, we found that the improvement remains approximately constant regardless of the maximum redshift considered.

This work is based on some modelling simplifications. Firstly, we did not take into account any uncertainties on observable-mass scaling relation and derived redshifts for the considered simulated tracers, as if galaxy cluster masses were perfectly measured. Also, in the building of the redshift-space catalogues, we mapped the tracer comoving coordinate into the observed ones by assuming the true
cosmological parameters of the simulation, taking into account only the geometrical distortions arising from the cosmology variation acting in the MCMC analysis. Moreover we modelled the geometric distortions for voids only, since these effects are expected to be negligible in the counts of galaxy clusters, the latter being less-spatially extended objects than voids. Our study was also limited
to the constraints on $\Omega_{\rm m}$ and $\sigma_8$ and it will be fundamental to consider also the remaining cosmological parameters on which the analysed probes strongly depend. Finally, we neglected the super-sample covariance (SSC) \citep{Hu&kravtsov03, Takada07, Takada14, Krause17, Lacasa18}, which concerns the fact that simulations and observations map a limited portion of the
Universe. Indeed, due to the missing modes larger than the simulation box, neglecting the SSC might introduce non-negligible effects. Recently, \citet{Bayer22} estimated the SSC specifically for cluster and void statistics. Although the SSC effect can be considered negligible for cluster and void counts alone, this is no longer true when considering their cross-correlation, potentially causing a change in the relative improvements of the combination analysis \citep{Takada07, Takada14, Lacasa16}. We will address all these issues in future works.

In order to obtain well grounded statistical results from the analysis of large-scale structures, it would be interesting to search for galaxy clusters and cosmic voids in much larger cosmological volumes. Upcoming galaxy surveys like the \textit{Euclid} mission \citep{Laureijs11, Scaramella14, Amendola18}, the Dark Energy Spectroscopic Instrument (DESI) \citep{DESI1, DESI2}  and Vera C. Rubin Observatory LSST \citep{LSST12} will map a very large fraction of our Universe, with volumes of the order of $10^2$ ($h^{-1}$ Gpc)$^3$, giving even more tight constraints on cosmological parameters and, mostly, improving our knowledge about the underlying cosmological model. The tighter constraints achieved by combining cluster and void counts, compared to those obtained by exploiting cluster counts alone, will significantly contribute to break cosmic degeneracies.

In this analysis we proved how the high complementarity of these cosmological probes, together with the great error reduction given by their combination, will provide a fundamental contribution in shedding light on tensions possibly affecting cosmological observations. We thus conclude that the methods presented in this work, including model calibrations on suitable mock catalogues, will motivate the exploitation of the combined analysis of galaxy cluster and cosmic void counts from future wide field galaxy surveys.

\section*{Acknowledgements}
We thank the referee for the useful suggestions, which helped us to improve the paper. We acknowledge the use of computational resources from the parallel computing cluster of the Open Physics Hub\footnote{\url{https://site.unibo.it/openphysicshub/en}} at the Physics and Astronomy Department in Bologna. CG and LM acknowledge the support from the grant PRIN-MIUR 2017 WSCC32 ZOOMING, and the support from the grant ASI n.2018-23-HH.0. CG acknowledges funding from the Italian National Institute of Astrophysics under the grant "Bando PRIN 2019", PI: Viola Allevato and from the HPC-Europa3 Transnational Access programme HPC17VDILO. KD acknowledges support by the COMPLEX project from the European Research Council (ERC) under the European Union’s Horizon 2020 research and innovation program grant agreement ERC-2019-AdG 882679 as well as support through the Deutsche Forschungsgemeinschaft (DFG, German Research Foundation) under Germany’s Excellence Strategy - EXC-2094 - 390783311. The calculations for the hydrodynamical simulations were carried out at the Leibniz Supercomputer Center (LRZ) under the project pr83li. 

\section*{Data Availability}
The simulation data underlying this article will be shared on reasonable request to the corresponding author.

%does not introduce relevant numerical biases on the final constraints.



\bibliographystyle{mnras}
\bibliography{main} % if your bibtex file is called example.bib

%\begin{thebibliography}{99}
%\end{thebibliography}

%%%%%%%%%%%%%%%%%%%%%%%%%%%%%%%%%%%%%%%%%%%%%%%%%%

%%%%%%%%%%%%%%%%% APPENDICES %%%%%%%%%%%%%%%%%%%%%
%%%%%%%%%%%%%%%%%%%%%%%%%%%%%%%%%%%%%%%%%%%%%%%%%%

\appendix

\section{Covariance matrix convergence}\label{appendix:a}

In this work, we derived cosmological constraints from hydrodynamical simulations by estimating the cross-covariance matrix, $\mathcal{C}$, between galaxy cluster and void number counts. It is important to verify that the noise in the covariance does not have a significant impact on the results, as it could potentially affect the estimation of the cosmological parameters in the Bayesian analysis. Therefore, in this Appendix we test the stability of our results using different sampling techniques.

In Section \ref{cross-cov} we computed $\mathcal{C}$ by dividing the tracer catalogues (galaxy clusters and voids) into $N_{\rm sub} = 125$ equally sized sub-regions. To test the covariance convergence and stability we compare now both jackknife and bootstrap resampling methods \citep{Efron82, Norberg08}. When adopting the former, we re-compute $\mathcal{C}$ as in Section \ref{cross-cov}, but varying this time $N_{\rm sub} \in [8,1000]$. Then, we extract again the combined constraints on $\Omega_{\rm m}$ and $\sigma_8$ for each considered case. When adopting the bootstrap technique, we  derive instead the cosmological constraints as a function of the number of resamplings $N_{\rm mock}$, considering a range of $N_{\rm mock} \in [100, 2400]$. For these tests we fixed the
nuisance model parameters of HMF and VSF (i.e. $a_1, a_2, a_3, A_1, A_2, B_\mathrm{slope}, B_\mathrm{off}$) to those calibrated in Section \ref{sec:calibrationHMF}, considering all data sets up to $z=1$.

\begin{figure}
	\includegraphics[width=1.1\columnwidth]{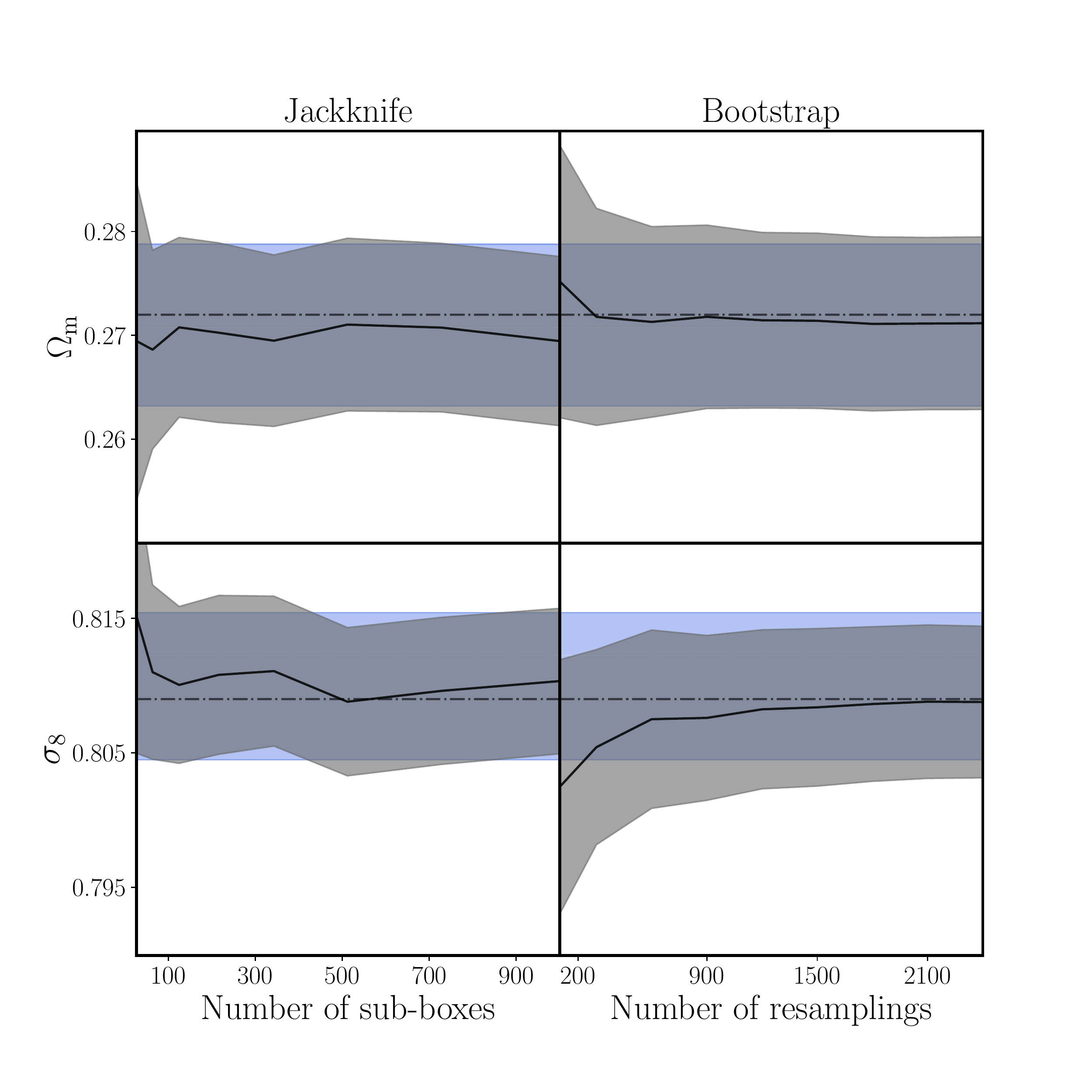}
    \caption{Posterior median values (black solid lines) for $\Omega_{\rm m}$ and $\sigma_8$ (top and bottom panels, respectively) as a function of the number of jackknife sub-boxes (\textit{left panels}) or the number of bootstrap resamplings (\textit{right panels}) used for the cross-covariance matrix estimation. The presented constraints are obtained considering all data sets up to $z = 1$. The shaded gray regions represent the $68\%$ confidence intervals around the posterior median of the presented cases. We report with blue shaded bands the $68\%$ confidence intervals of the constraints obtained by applying the jackknife sampling technique with $N_\mathrm{sub}=125$ (see Section \ref{test_constraints}). The dotted-dashed black lines represent the true cosmological values of the \textit{Magneticum} simulations.}
    \label{cov_conv}
\end{figure}

In Figure \ref{cov_conv} we report the median values for $\Omega_{\rm m}$ and $\sigma_8$ and the corresponding $68\%$ uncertainties obtained for the different cases analysed. We note that the cosmological constraints are stable and consistent with those reported in Eq. \eqref{final_constr}, for all the methodologies employed. We conclude that the final constraints presented in this work, obtained by adopting the jackknife sampling technique with $N_\mathrm{sub}=125$, can be considered unaffected by relevant numerical issues related to the estimation of the covariance matrix.

% Don't change these lines
\bsp	% typesetting comment
\label{lastpage}
\end{document}